\documentclass[
 reprint,
preprintnumbers,
 amsmath,amssymb,
 aps,
prd,
nofootinbib
]{revtex4-2}

\usepackage[normalem]{ulem} 
\usepackage{graphicx}  
\usepackage{color}
\usepackage{dcolumn}   
\usepackage{bm}        
\usepackage{hyperref}  
\usepackage{amsmath, amssymb} 
\usepackage{physics}
\usepackage{float}
\usepackage{comment}

\graphicspath{
{./}
  {./media/}
  {./media/fit_plots/}
  {./media/misc_plots/}
  {./media/velocity_plots/}
  {./media/xi_plots/}
}

\def\al{\alpha}

\def\de{\delta}

\def\la{\lambda}
\def\rh{\rho}
\def\si{\sigma}
\def\ta{\tau}

\def\Ga{\Gamma}

\def\La{\Lambda}

\def\pa{\partial}

\newcommand{\ben}{\begin{equation}}
\newcommand{\een}{\end{equation}}
\newcommand{\bea}{\begin{eqnarray}}
\newcommand{\eea}{\end{eqnarray}}
\newcommand{\ba}{\begin{array}}
\newcommand{\ea}{\end{array}}
\newcommand{\bal}{\begin{align}}
\newcommand{\eal}{\end{align}}
\newcommand{\bit}{\begin{itemize}}
\newcommand{\eit}{\end{itemize}}

\newcommand{\bx}{\textbf{x}}
\newcommand{\bX}{\textbf{X}}

\newcommand{\units}[1]{\,\text{#1}}

\newcommand{\partialm}{{\partial_{\mu}}}

\newcommand{\nn}{\nonumber \\}
\newcommand{\alm}{\alpha_\text{m}}

\newcommand{\gm}{g_\text{m}}

\newcommand{\mMon}{M_\text{m}}
\newcommand{\Mpl}{M_\text{Pl}}

\newcommand{\nm}{n_\text{m}}

\newcommand{\Omm}{\Omega_\text{m}}

\newcommand{\rhom}{\rho_\text{m}}

\newcommand{\rMon}{r_\text{m}}

\newcommand{\tauCG}{\tau_\text{cg}}

\newcommand{\Ti}{T_\text{i}}
\newcommand{\Tf}{T_\text{f}}

\newcommand{\Ym}{Y_\text{m}}

\newcommand{\Zeff}{Z_\text{eff}}

\begin{document}

\preprint{HIP-2025-31/TH}

\title{Numerical simulations of magnetic monopole evolution in an expanding universe}

\newcommand{\Sussex}{\affiliation{
Department of Physics and Astronomy,
University of Sussex, Falmer, Brighton BN1 9QH,
U.K.}}

\newcommand{\HIPetc}{\affiliation{
Department of Physics and Helsinki Institute of Physics,
PL 64, 
FI-00014 University of Helsinki,
Finland
}}

\newcommand{\UPV}{\affiliation{
Department of Physics, University of the Basque Country UPV/EHU, Bilbao, Spain
}}

\newcommand{\QC}{\affiliation{
EHU Quantum Center, University of the Basque Country UPV/EHU, Bilbao, Spain
}}

\author{Mark Hindmarsh}
\email{mark.hindmarsh@helsinki.fi}
\HIPetc
\Sussex

\author{Asier Lopez-Eiguren}
\email{asier.lopez@ehu.eus}
\UPV
\QC

\author{Riikka Seppä}
\email{riikka.seppa@helsinki.fi}
\HIPetc

\author{David J. Weir}
\email{david.weir@helsinki.fi}
\HIPetc

\begin{abstract}
   
   Magnetic monopoles are an inevitable feature of post-inflation symmetry-breaking phase transitions in grand unified theories. Analytic estimates of their density indicate that they are compatible with standard cosmology only if their mass is less than $10^{11}$ GeV \cite{Zeldovich:1978wj,Preskill:1979zi}. We initiate a programme of numerical studies of monopole dynamics by simulating a gas of 't Hooft-Polyakov monopoles formed by the Kibble mechanism after a phase transition. In this paper we simulate monopoles in a radiation background, but without interactions with the radiation, in order to resolve differences between analytical models. We find that during the radiation era, the monopoles find each other and annihilate efficiently enough to keep their density fraction constant, which supports the modelling of Zel'dovich and Khlopov \cite{Zeldovich:1978wj}  and Preskill \cite{Preskill:1979zi} in the epoch when plasma interactions can be neglected. In the matter era the density fraction decreases logarithmically. Further work is needed to quantify the  effect of the thermal bath, which is expected to reduce the annihilation rate at later times.
\end{abstract}

\maketitle

\section{Introduction}

The early universe
is thought to have gone through a number of symmetry-breaking phase transitions. When the vacuum manifold after a phase transition is topologically non-trivial, 
topological defects may form~\cite{Kibble:1976sj}. These are nontrivial, extended field theory solutions 
including domain walls, cosmic strings, monopoles and textures (see e.g.~\cite{Vilenkin:2000jqa,Manton:2004tk,Vachaspati:2006zz}).

A gauged, 't Hooft-Polyakov monopole~\cite{Hooft:1974qc,Polyakov:1974ek} can form when 2-spheres embedded in the vacuum manifold cannot all be contracted to a point - for example, when the symmetry-breaking is $\mathrm{SU}(2) \to \mathrm{U}(1)$ the vacuum is itself a 2-sphere. In general, the vacua of any grand unified theory (GUT) undergoing symmetry breaking will admit 't Hooft-Polyakov monopole solutions~\cite{Hooft:1974qc,Polyakov:1974ek,Goddard:1981rv}.

The magnetic charge of a monopole $\gm$ obeys the Dirac quantisation condition $q\gm = 2n\pi$, where $n$ is an integer and $q$ is the electric charge of any ordinary particle~\cite{Dirac:1931kp}. Taking $n=1$ and setting $q$ to the fundamental electric charge $e$, the magnetic charge $g_D = 2\pi/e$ is termed the Dirac charge. Searches for magnetic monopoles often frame their constraints in terms of integer multiples of $g_D$.

Monopoles produced by a GUT phase transition 
would act as a massive particle species and dominate the energy density of the Universe at unacceptably early times if their mass is greater than around $10^{11}$ GeV  \cite{Preskill:1979zi,Zeldovich:1978wj}, constituting the `monopole problem' of grand unification. Theories of inflation eliminate this problem by diluting the cosmological energy density of monopoles into insignificance \cite{Guth:1980zm}. It is, however, still possible that monopoles could have been produced after inflation, 
at a lower-temperature phase transition unconnected with grand unification. 
Then, if monopoles are more massive than around $10^{11}$ GeV, some other mechanism must act to reduce their density. Proposals include an intermediate phase of electromagnetic symmetry-breaking~\cite{Langacker:1980kd,Hikasa:1981zg} and monopole dissociation on domain walls~\cite{Dvali:1997sa,Pogosian:1999zi,Brush:2015vda,Bachmaier:2023zmq}.

Astrophysical processes also place constraints on monopole flux at the Earth. Magnetic monopoles would be accelerated along magnetic field lines associated with galaxies, which in turn would dissipate the galactic magnetic field. If the magnetic field is continuously regenerated by dynamo action, an upper bound -- the Parker bound -- is placed on the abundance of monopoles, above which the magnetic field would be dissipated by the action of the monopoles~\cite{Parker:1970xv,Turner:1982ag}.  For monopoles with mass $\mMon \lesssim 10^{17}$ GeV, the Parker bound is 
$F < 10^{-15} \, \text{cm}^{-2} \, \text{sr}^{-1} \, \text{s}^{-1} . $
This can be recast as a bound on the monopole density fraction today for monopoles in the mass range $10^{11} \lesssim (\mMon/\text{GeV})  \lesssim 10^{17}$,
\begin{equation}
\Omm \lesssim \phi_g^{-1} 10 \left( \frac{\mMon}{10^{17}\,\text{GeV}} \right)^{3/2},
\end{equation}
where $ \phi_g$ is the density enhancement of monopoles in the galaxy. Stronger bounds can be obtained with assumptions about seed magnetic fields at the epoch of galaxy formation \cite{Adams:1993fj} and in the early universe \cite{Kobayashi:2022qpl}.  If monopoles catalyse proton decay, then the constraints are even stronger (see e.g. \cite{ParticleDataGroup:2024cfk}).

Other constraints on the physics of monopoles come from collider experiments. Pairs of 
magnetic monopoles, if sufficiently light, could be created at colliders. 
The Monopole and Exotics Detector Experiment at the LHC (MoEDAL) has been designed to directly search for them, 
and has placed constraints on their mass.
For proton interactions~\cite{MoEDAL:2023ost},
considering monopole charges ranging from 1$g_D$ to 10$g_D$, monopoles with masses below 3.9~TeV are excluded and in heavy-ion Pb+Pb interactions~\cite{MoEDAL:2024wbc}, 
monopoles between 2$g_D$ and 45$g_D$, with masses below 80~GeV are also excluded. Other LHC experiments have also placed constraints on magnetic monopole production. Again using Pb+Pb collision data, the ATLAS experiment has ruled out  
monopoles with charge 1$g_D$ and mass below 120~GeV~\cite{ATLAS:2024nzp}.

Hence, if monopoles exist as possible states, they must be both rare and massive. In this paper we address the problem of calculating their number density, assuming that they are produced in an early universe phase transition after inflation \cite{Kibble:1976sj,Rajantie:2002dw}. 

So far, only analytic modelling of gauge monopole gases has been carried out \cite{Zeldovich:1978wj,Preskill:1979zi,Martins:2008zz,Sousa:2017wvx}, although there have been numerical simulations of systems of necklaces (monopoles trapped on strings) \cite{Hindmarsh:2016dha, Hindmarsh:2018zch} 
and global monopoles \cite{Yamaguchi:2001xn,Lopez-Eiguren:2016jsy}.  Both necklaces and global monopoles exhibit scaling, meaning that their mean separation increases in proportion to the causal horizon of the Universe, and therefore their density fraction remains constant in the case of necklaces, or drops inversely with scale factor in the case of global monopoles. However, necklaces are only a potential solution to the gauge monopole problem if there is intermediate-scale electromagnetic symmetry breaking.

In Ref.~\cite{Dvali:1997sa}, an alternative solution to the monopole problem was proposed, termed `defect erasure'. In models possessing more than one class of defect solution with different dimensionality, a lower-dimensional defect can be `erased' by interacting with a higher-dimensional defect. The prototypical example is of 't Hooft-Polyakov monopoles getting erased by domain walls. Upon encountering a domain wall, the monopole unwinds; the domain wall network itself can then efficiently annihilate before the present day. This phenomenon has been well-studied numerically in interactions between individual defects~\cite{Pogosian:1999zi,Brush:2015vda,Bachmaier:2023zmq}.

In this paper we initiate the numerical study of the evolution of a gas of monopoles and antimonopoles, in order to improve analytic modelling of their evolution.  In particular, we seek to resolve differences between Refs.~\cite{Martins:2008zz} and \cite{Sousa:2017wvx} over the modelling of the average acceleration of monopoles towards antimonopoles. 
To do so, we need only to simulate the case of monopoles which do not interact with the radiation plasma dominating the energy density of the universe, which would add a significant layer of complexity.

The models of Refs.~\cite{Martins:2008zz} and \cite{Sousa:2017wvx} both propose modifications of 
the Coulomb law interaction between monopoles and antimonopoles an expanding universe, in the form of a multiplicative power-law of the ratio 
of the mean separation to the horizon distance, although they differ in what the index of the power law should be.
We find that our results are consistent with there being no modification, in the sense that they produce a clear fixed point in the evolution of the dimensionless mean separation and velocity variables of the model.

While we cannot constrain the index of the power law well, we find that the proposal of Ref.~\cite{Martins:2008zz} does not produce a clear fixed point in the evolution of the dimensionless mean separation and velocity variables of the model. The range of power laws proposed in Ref.~\cite{Sousa:2017wvx} produce progressively less clear fixed points as the parameter departs from zero (no modification of the Coulomb law).
Further constraining the latter model is beyond the scope of the current work.

We also study the density fraction of monopoles. In the radiation era simulations, starting from a wide range of initial values, the density fractions evolve towards a value of around $\Omega_{\rm m} \simeq  0.1 GM_{\rm m}^2$.  A constant density fraction is in agreement with \cite{Zeldovich:1978wj,Preskill:1979zi,Hill:1982iq} when plasma interactions can be neglected.

In Section~\ref{sec:expectations} we discuss theoretical modelling of networks of monopoles in the early universe, in Section~\ref{sec:model} we give the details of our model and simulation. Our measurements are described in Section~\ref{sec:measure}, and the results of our simulations in Section~\ref{sec:res}. We discuss our findings and conclude in Section~\ref{sec:dis}.

\section{Modelling a monopole gas}
\label{sec:expectations}

\subsection{Review}

The starting point is the rate of change of monopole density $\nm$. The monopole density is diluted by the expansion of the universe, and by annihilation.  Assuming that annihilation proceeds by random encounters with velocity-averaged cross-section $D$, the evolution of the number density $n_\mathrm{m}$ can be written~\cite{Zeldovich:1978wj,Preskill:1979zi}
\begin{equation}
   \label{eq:Preskilldiffusion}
   \dot{n}_\mathrm{m} + 3H n_\mathrm{m} = -  D n_\mathrm{m}^2,
\end{equation}
However, the assumption that monopoles are freely moving particles is open to question: they have strong Coulomb interactions with each other, and they are strongly coupled to the conducting plasma of the early universe. The forces of attraction and friction must therefore be taken into account.

Before examining the details, one might guess that the temperature dependence of $D$ has the form~\cite{Preskill:1979zi} 
\begin{equation}
D =  \frac{A}{\mMon^2} \left(\frac{\mMon}{T}\right)^p,
\end{equation}
where  
$T$ is the temperature, $A$ is a dimensionless constant and $p$ is a parameter encoding the annihilation dynamics.  Using the Friedmann equation $H = C T^2/\Mpl$, where $\Mpl$ is the Planck mass and $C$ is a constant, Eq.~\eqref{eq:Preskilldiffusion} can be integrated to give
\begin{equation}
\Ym^{-1}(T) = \Ym^{-1}(\Ti) + \frac{AC}{p - 1} \frac{\Mpl}{\mMon} \left[\left(\frac{\mMon}{T}\right)^{p-1} - \left(\frac{\mMon}{\Ti}\right)^{p-1} \right] ,
\end{equation}
where $\Ym = \nm/s$ is the monopole yield and $s \propto T^3$ is the entropy density. In the case $ p > 1$, the yield decreases asymptotically as 
\begin{equation}
\Ym(T) \to \frac{AC}{p-1} \frac{\mMon}{\Mpl} \left(\frac{T}{\mMon}\right)^{p-1},
\end{equation}
while for $p<1$ there is no substantial reduction in the yield.

It was argued \cite{Zeldovich:1978wj,Preskill:1979zi} that well below the phase transition forming the monopoles, 
the annihilation rate is limited by the rate at which monopoles diffuse towards each other.  
Once a monopole-antimonopole pair is separated by less then a distance $a_c = \gm^2/4\pi T$, where the binding energy equals the thermal energy, it will inevitably annihilate. The effective cross section is therefore $4\pi a_c^2$.  The inward speed at this radius is determined by the balance between Coulomb attraction and friction, $v_c \simeq (\tau_\text{sc}/m) \gm^2/4\pi a_c^2$,
where $\tau_\text{sc}$ is the mean free time between large angle scatterings.  Hence the effective velocity averaged cross-section is $D \simeq  \gm^2(\tau_\text{sc}/m) $.   The scattering rate of massless particles in the plasma, which have density $n \sim T^3$ and momentum $T$, is $n \si $, where $\si \sim q^2/T^2$.  Each scattering event changes the momentum of the monopole by a fraction $T/m$, and hence the mean free time of the monopoles is $\tau_\text{sc} \sim m/T^2$.  Therefore, in the diffusion-limited annihilation case, we have $p=2$, and monopoles annihilate efficiently.

At lower temperatures, the monopole mean free path exceeds the radius at which the thermal energy equals the binding energy. The annihilation rate is then governed by the Coulombic dynamics of a monopole-antimonopole pair.   
The picture in Refs.~\cite{Zeldovich:1978wj,Preskill:1979zi} is that monopoles approach one another with a thermal velocity $v \simeq \sqrt{T/\mMon}$, and the excess kinetic energy must be radiated away in order for the pair to become bound, and eventually annihilate.  The cross-section for this process was estimated as $\si_C \simeq \pi a_c^2(T) (T / \mMon)^{6/10}$.  Hence the effective velocity-averaged cross-section for Coulomb annihilation has parametric dependence 
\begin{equation}
D_C \sim (\gm^4/\mMon^{2})(T/\mMon)^{-9/10}.
\end{equation}
We therefore have $p < 1$ and annihilation effectively stops. This happens at a temperature $\Tf \sim \mMon/\gm^4$, and the monopole yield is frozen at 
\begin{equation}
\Ym(\Tf)  \sim \frac{1}{\gm^6} \frac{\mMon}{\Mpl}  .
\end{equation} 
The energy density of the monopoles grows as $(\mMon/T)\Ym(\Tf)$ in the radiation era, and  dominates before matter-radiation equality, unless $\mMon \ll \gm^3 10^5 \units{GeV}$. As the phase transition producing monopoles is supposed to be breaking grand unified symmetries, this leads to the famous monopole problem.

An alternative modelling framework is the velocity one-scale (VOS) model, which is used to describe the large-scale motion of networks of extended topological defects \cite{Martins:1996jp}. VOS models describe the evolution of the characteristic length scale of the defect network and the root-mean-square (RMS) velocity. VOS models have also been applied to monopoles \cite{Martins:2008zz,Sousa:2017wvx}, where the characteristic length scale $L$ has been defined in terms of the energy density by $L^3 = \mMon/\rhom$. This expression is valid in the non-relativistic limit $v \ll 1$ where the kinetic energy of the monopoles can be neglected in comparison to the mass.

Eq.~\eqref{eq:Preskilldiffusion} then becomes
\begin{align}
   \frac{\mathrm{d}L}{\mathrm{d} t} & = HL  + \frac{1}{3}\frac{D}{L^2},
   \label{e:OldVOSL}
\end{align}
which can also be derived by energy conservation. 
From Newton's second law, the RMS velocity is held to obey
\begin{align}
   \frac{\mathrm{d} v}{\mathrm{d}t} & = - \left(H + \frac{1}{t_\text{f}}\right)v + \frac{k}{\mMon L^2} \frac{1}{(HL)^\alpha} .
    \label{eq:ma_vos_vel} 
\end{align}
The first term on the right hand side models damping effects: $t_\text{f}$ is a time scale associated with friction. The second term models the magnetic Coulomb force due to the other monopoles within the cosmological horizon: $k$ is a dimensionless force parameter (expected to be proportional to $\gm^2$) and the index $\alpha$ is proposed to take into account correlations between monopoles which alter the distance dependence of the Coulomb attraction.Note that $(HL)^{-3}$ is the number of monopoles within the cosmological horizon.

However, the exact value of the index $\alpha$ is the subject of some disagreement: Ref.~\cite{Martins:2008zz} argued for $\alpha = -1$. On the other hand, 
Ref.~\cite{Sousa:2017wvx} 
argued for $\alpha \geq 0$. We will see that our simulations are consistent with $\alpha = 0$.

\subsection{Modelling the comoving monopole separation $\xi$}
\label{sec:modelling_xi}

We propose a slightly different formulation of the VOS model, and apply it to the case where thermal effects can be neglected, relevant for our simulations. We suppose that the correlations between monopoles with like-sign charges and opposite sign charges are similar, and that a monopole easily 
finds and annihilate with the nearest antimonopole, drawn together by the Coulomb force. 
A monopole-antimonopole pair then has average separation of order $L$, and if they have RMS velocity $v$, they will take a time of order $v/L$ to find each other and then annihilate.   
This suggests an equation for the monopole density 
\begin{equation}
\frac{\mathrm{d} \nm }{\mathrm{d} t}   + 3 H \nm = - d \frac{v}{L} \nm,
\label{e:OurNmEq}
\end{equation}
where $d$ is a constant. This is apparently different in form from Eq.~\eqref{eq:Preskilldiffusion}, but they can be reconciled if one supposes that the effective velocity-averaged annihilation cross-section is $D = d L^2/ v$.

Translating Eq.~\eqref{e:OurNmEq} into an equation for $L$, 
\begin{equation}
\frac{\mathrm{d}L}{\mathrm{d} t}  =  H {L} + \frac{d}{3} v,
\end{equation}
and the velocity is determined through the dynamical equation 
\begin{equation}
\frac{dv}{dt} = -Hv +  \frac{\alm}{\mMon} \frac{1}{L^2}
\end{equation}
where $\alpha_\text{m} = \gm^2/4\pi$. 
The equations are equivalent to Eqs.~(\ref{e:OldVOSL}, \ref{eq:ma_vos_vel}), with $k = \alm$ and 
$\alpha=0$.
They have the solution $L \propto t^{2/3}$, $v \propto t^{-1/3}$.

Our simulations use comoving coordinates, so it is useful to 
have expressions in terms of comoving distance $\xi = L/a$ and conformal time $\tau$. The velocity expressed in comoving displacement and conformal time is  
equal to the physical peculiar velocity. 
Hence we have 
\begin{eqnarray}
\frac{d\xi}{d\tau} &=& \frac{d}{3} v,  \label{e:NewVOSxi}\\
\frac{dv}{d\tau} &=&   - {\cal H} v + \frac{\alm}{\mMon} \frac{1}{a\xi^2} . \label{e:NewVOSv}
\end{eqnarray}
where ${\cal H} = d\ln(a)/d\tau = \nu/\tau$, with $\nu = 1$ in the radiation era and $\nu = 2$ in the matter era. A more careful derivation of Eq.~\eqref{e:NewVOSv} is given in Appendix~\ref{app:mono_dynamics}.

The equations have an explicit length scale
\begin{equation}
\rMon = \alm/\mMon .
\end{equation}
We make the substitutions
\begin{eqnarray}
\xi &=& x \tau \left( \frac{\rMon }{a(\tau) \tau} \right)^{1/3}, \\
v &=& y  \left( \frac{\rMon}{a(\tau) \tau} \right)^{1/3}.
\end{eqnarray}

Then
\begin{eqnarray}
\tau \dot{x} &=& \frac{1}{3} (y d - (2-\nu) x), \\
\tau \dot{y} &=&  \frac{\Zeff^2}{x^2} - \frac{2- \nu}{3} y .
\end{eqnarray}
In the radiation era, where $\nu = 1$, these equations have a fixed point at 
\begin{equation}
x_* = (3 \Zeff^2 d)^{1/3},\quad 
y_* = \left(3 \frac{\Zeff^2}{ d^2}\right)^{1/3} .
\end{equation}
Expanding around the fixed point, $(x_* + \de_x, y_* + \de_y)$, the approach is described by the equation
\begin{equation}
\frac{d}{d\La}
\left(\begin{array}{c}
\de_x \\[3pt]
\de_y
\end{array}\right)
=
\left(\begin{array}{cc}
 - \frac{1}{3} & \frac{1}{3} d \\[3pt]
- \frac{2}{3\Zeff^2 d}  & - \frac{1}{3} 
\end{array}\right)
\left(\begin{array}{c}
\de_x \\[3pt]
\de_y
\end{array}\right) ,
\end{equation}
where $\La = \ln(\tau/\tau_0)$. 
The matrix has a pair of complex eigenvalues,
\begin{equation}
\la_\pm = -\frac{1}{3} \pm i \sqrt{\frac{2}{9\Zeff^2}} .
\end{equation}
The real part of both eigenvalues is negative, and hence the fixed point is attractive. 

In the matter era, where ${\cal H}  = 2/\tau$, the equations become
\begin{eqnarray}
\tau \dot{x} &=& \frac{1}{3} y d , \\
\tau \dot{y} &=&  \frac{\Zeff^2}{x^2} .
\end{eqnarray}
In this case, there is a line of fixed points at $x_* = \infty$, with any value of $y_*$.  
Writing $\La = \ln(\ta/\ta_0)$, the fixed point is approached as
\begin{equation}
x(\La) \sim  \frac{1}{3} (y_* d) \La, \quad
y(\La) \sim y_* - \left( \frac{3\Zeff}{y_* d} \right)^2 \frac{1}{\La} . 
\end{equation}
The monopole density fraction is 
\begin{equation}
\Omm \equiv \frac{\rhom}{\rho_\text{c}} = \frac{8\pi G\mMon^2}{3 \nu^2 \alm } \frac{1}{x^3},
\label{eq:densityfraction}
\end{equation}
where $\rho_\text{c} = 3H^2/8\pi G$ is the critical density. 
Hence, in the radiation era, the model predicts that the monopole density fraction tends to a constant (with $\xi \sim \tau^{1/3}$ i.e. $L \sim t^{2/3}$), while in the matter era, it predicts that the density fraction decreases as $1/\ln^3(\tau/\tau_0)$.  

The mean monopole separation is always less than the causal horizon distance, as it is controlled by the ratio of the microscopic scale $\rMon$ and $a(\tau)\tau$, which is proportional to the horizon distance $d_\text{h} = a(\tau)\tau/(\nu+1)$. Taking the ratio of the comoving separation and the comoving horizon distance, 
\begin{equation}
\frac{\xi}{\tau} = x  \left(\frac{\rMon}{a(\tau)\tau}\right)^{1/3} \ll 1 .
\end{equation}
Hence the horizon distance would not seem to be a significant length scale in the dynamics of a monopole gas in an expanding universe. Nonetheless, we can explore the effect of altering the Coulomb attraction term, along the lines of Eq.~\ref{e:OldVOSL} as suggested in Ref.~\cite{Martins:2008zz}, to 
\begin{equation}
\frac{dv}{d\tau} =   - {\cal H} v +  \Zeff^2\frac{\alm}{\mMon} \frac{1}{a\xi^2}\left(\frac{\tau}{\xi}\right)^\al,
\end{equation}
where we have absorbed a factor $\nu^{\al}$ in the effective charge squared $\Zeff^2$.

In this case we can make the substitutions
\begin{eqnarray}
\xi &=& x_\al \tau \left( \frac{\rMon }{a(\tau) \tau} \right)^{1/(3+\al)}, \\
v &=& y_\al  \left( \frac{\rMon}{a(\tau) \tau} \right)^{1/(3+\al)},
\end{eqnarray}
to find
\begin{eqnarray}
\tau \dot{x}_\al &=& \frac{1}{3} y d - \frac{2-\nu}{3+\al} x, \\
\tau \dot{y}_\al &=&  \frac{\Zeff^2}{x^{2+\al}} - \frac{2- \nu}{3+\al} y .
\end{eqnarray}
Hence the new variables $x_\al$ and $y_\al$ also have a fixed point, but they translate back to a different prediction for the time dependence of the mean comoving separation $\xi$ and the RMS velocity $v$.

\section{Model and simulation}
\label{sec:model}

In comoving coordinates $x^i$, conformal time $\tau = x^0$, and with
scale factor $a$, the action is
\begin{eqnarray}
\mathcal{S} = \int d^4x \Bigg( -\frac14 F^a_{\mu\nu} F^{\mu\nu a}  + a^2 \tr (D_{\mu} \Phi D^{\mu}\Phi) - a^4 V(\Phi)  \Bigg), \nonumber
\end{eqnarray}
where $D_{\mu}=\partialm+igA_{\mu}$ is the covariant derivative, and
$A_{\mu}=A_{\mu}^a\si^a/2$, where $\sigma^a$ are Pauli matrices. The Higgs field is in adjoint representation, $\Phi = \frac12 \phi^a\sigma^a$. Spacetime indices have been raised with the
Minkowski metric with mostly negative signature.

The potential can be written as
\begin{eqnarray}
V(\Phi) &=& - m^2\tr \Phi^2 + \lambda (\tr \Phi^2)^2.
\label{eq:potential}
\end{eqnarray}
The vacuum expectation value is then $\tr \Phi^2 = \eta^2$, where $\eta^2 = |m^2|/\lambda$. The system undergoes a symmetry breaking phase transition $\mathrm{SU}(2)\rightarrow \mathrm{U}(1)$. In the broken phase the system has 't Hooft-Polyakov monopole solutions with mass~\cite{Forgacs:2005vx} 
\begin{equation}
\label{eq:monopolemass}
\mMon = \frac{4\pi \eta}{g}f\biggr(\frac{m_\mathrm{H}}{m_\mathrm{W}}\biggr), \quad f(1) \approx 1.238.
\end{equation}
The Higgs particles then have mass $m_\mathrm{H} \equiv \sqrt{2} m$, and in addition to a massless photon there are two massive charged gauge bosons with masses $m_\mathrm{W} \equiv g\eta = gm/\sqrt{\lambda}$. One typically bases the comoving monopole width on the scalar mass, $w_\mathrm{m} \equiv 1/(a m_\mathrm{H})$, but it is possible to define a size based on the gauge boson masses as well, $w_\mathrm{m,g} \equiv 1/(a m_\mathrm{W})$.

\subsection{Simulating a monopole network in an expanding universe}

The dynamical range of the simulations is constrained by the shrinking of the monopoles when compared to the growing lattice spacing. To combat this, we use the extension of the Press-Ryden-Spergel technique~\cite{Press:1989yh} to gauge theories \cite{Bevis:2010gj} to grow the monopole width before the physical evolution.

Couplings and mass parameters are scaled with $a^{1-s}$, where $a$ is the scale factor and $0 \le s \le 1$. Specifically, we have\footnote{Note that in Ref.~\cite{Hindmarsh:2016dha} the scaling of the gauge coupling erroneously mentions $g$ rather than $g^2 \to g^2/a^{2(1-s)}$ in Eq.~(A2). This incorrect rescaling is in any case inconsistent with the other equations in the Appendix of that paper.}
\begin{equation}
   m^2 = \frac{m_0^2}{a^{2(1-s)}}, \quad \lambda = \frac{\lambda_0}{a^{2(1-s)}}, \quad \text{and} \quad g^2 = \frac{g_0^2}{a^{2(1-s)}}, \label{eq:rescaled_couplings}
\end{equation}
where $m_0$, $\lambda_0$ and $g_0$ are constants. This method keeps the vacuum expectation value of the Higgs field fixed. The ratio of the masses of the scalar and massive gauge bosons also remains fixed, and hence the static monopole solution remains exactly the same~\cite{Forgacs:2005vx}. In other words, the ratio of the monopole widths measured by the scalar or gauge fields remains the same.

\begin{figure} [t]
   \centering \includegraphics[width=0.45\textwidth]{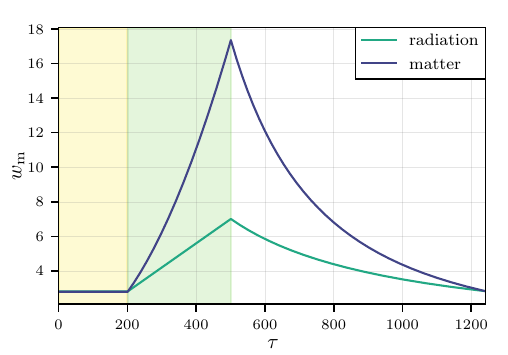}
   \caption{Comoving size of the monopoles in our simulations, relative to the lattice spacing $\mathrm{d}x=1$. For $\tau < 200$, the evolution is strongly damped. For $200 < \tau < 501$, Friedmann evolution starts, but couplings are scaled with a power of the scale factor so that the comoving monopole size can grow to its physical value.  Physical evolution starts at $\tau = 501$.
   }
   \label{fig:monosize}
\end{figure}

The comoving monopole width is modified to $w_m \sim (a^sm)^{-1}$. For $s=-1$ the physical monopole size grows linearly with the scale factor. If we do this for a time at the start of a simulation (a period of so-called `core growth'), then we can subsequently run for longer with the physical choice $s=1$ before the defects become too small to resolve on our lattice. In Figure~\ref{fig:monosize} we show how this helps to keep the monopole size larger than the comoving lattice spacing.

 \subsection{Lattice implementation}

We discretise the system on a three-dimensional comoving lattice with a cosmological background. The discretised Hamiltonian of this model is then, with temporal gauge $A_0 = 0$,
\begin{equation}
H(t) \equiv \sum_x a^2 T_{00}(x,t),
\end{equation}
where
\begin{align}
T_{00}(x,t)   =  & \frac{1}{2g_0^2a^{2s}} \sum_{i,a} \epsilon_i^a(x,t)^2 + \frac12 \sum_{a} \pi^a(x,t)^2 \nn
& \qquad + \frac{4}{g_0^2 a^{2s}} \sum_{i<j} \Bigl[ 1- \frac12 \Tr U_{ij}(x)  \Bigr] \nn
& \qquad -\sum_{i} 2\Tr \Phi(x)U_i(x)\Phi(x+\hat{\imath}) \nn
& \qquad + 6 \Tr\Phi^2 + a^2 V(\Phi),
\end{align}
in which we have used the rescaled parameters~(\ref{eq:rescaled_couplings}).

The link matrices are $U_{\mu} = u_{\mu}^0 + i \sigma^au_{\mu}^a$, with $(u_{\mu}^0)^2 + u_{\mu}^a u_{\mu}^a = 1$, with
\begin{align}
\epsilon_i^a = -\frac{i}{2} \Tr(\sigma^a\dot{U}_iU_i^{\dagger}).
\end{align}

The potential is 
\begin{align}
V(\Phi) = \frac{1}{a^{2(1-s)}}[-m_0^2 \Tr \Phi^2 + \lambda_0(\Tr \Phi^2)^2 ].
\end{align}

The equations of motion derived from the lattice action are evolved with standard leapfrog method. For the damping terms, the Crank-Nicolson method is used. More details in Ref.~\cite{Hindmarsh:2018zch}.

\subsection{Numerical Setup}

The SU(2) gauge fields are set to a cold configuration: the field to unity and the momentum to zero. The Higgs field is initialised with hot Gaussian values with the expectation normalised to the vacuum expectation value. 
We can control the correlation length of the Gaussian values which in turn approximately controls the amount of monopoles in the system initially. 

To do this, we work in momentum space for the components $\phi_n$ of $\Phi$. Denoting the momentum $\mathbf{k}$ associated with the $(n_x,n_y,n_z)$'th reciprocal lattice site as
\begin{equation}
   \mathbf{k} = \frac{2\pi}{L}(n_x, n_y, n_z); \quad -\frac{L}{2} < \{n_x,n_y,n_z\} \leq \frac{L}{2}
\end{equation}
we initialise
\begin{equation}
   \phi^a(\mathbf{x}) = \frac{1}{L^3} \sum_\mathbf{k} \tilde{\phi}^a(\mathbf{k})e^{i\mathbf{k}\cdot \mathbf{x}},
\end{equation}
where the real and imaginary parts of $\tilde{\phi}^a(\mathbf{k})$ are Gaussian random numbers with zero mean and variance determined by the power spectrum 
\begin{equation}
P(k) =  \eta^2 \pi \left(\dfrac{l_0}{L}\right)^3  \exp(-\frac{1}{2}k^2 l_0^2 ).
\end{equation}
In continuum formulation, the fields are initialised with
\begin{align}
\langle \tilde{\phi}^a(\mathbf{k}) \tilde{\phi}^b(\mathbf{q}) \rangle = (2\pi)^3 \delta_{ab} \delta^{(3)}(\mathbf{k} - \mathbf{q}) P(k).
\end{align}

We impose $\tilde{\phi}^a(\mathbf{k}) =  \tilde{\phi}^{a*}(-\mathbf{k})$ so that the position-space result is pure real. After the initialisation, we normalise the amplitude of the fields to the vacuum expectation value $\Phi^2 = \eta^2$.

After the initialisation, the fields are damped for a period of $200~\tau$ without expansion (see Appendix A of Ref.~\cite{Hindmarsh:2016dha} for more details; we take a damping factor of $\sigma = 4$ for both fields). This is done to remove unphysical excitations from the initialisation and to give the Higgs field time to settle down into monopole configurations. 

After the damping, a growth phase is performed to expand the dynamical range of the run. During this phase we set $s = -1$, so the monopole width grows. The growth phase length is set so that the monopole width at the beginning of the growth phase and at the end of the main simulation phase match (see Fig.~\ref{fig:monosize}). The length of the main simulation phase, where $s = 1$, is chosen so that growth phase and main simulation phase amount to the half-light crossing time.

We run simulations both in the radiation era, where $a \propto \tau$, and in the matter era, where $a \propto \tau^2$.  
For all simulations, the number of lattice sites is $L^3 = 2048^3$, the time step is $d\tau = 0.1$, the lattice spacing is $dx = 1$, and the mass and coupling parameters are $m^2 = 0.125$, $\lambda = 0.4$, $g = 1$.

Values of other parameters can be seen from Table \ref{tab:param}. We explore cases with four different initial separation parameters, only one of which we use in the matter era simulations. For each set of parameters, we perform 3-5 simulations.

\begin{table}[h]
\centering
\begin{ruledtabular}
\begin{tabular}{cccccc}
\multicolumn{6}{c}{Radiation} \\
$l_0$ & $n_{\text{sim}}$ & $\tau_\text{damp}$ & $\tau_\text{cg}$ & $\tau_{\text{end}}$ & $\xi_N(\tau = 0)$ \\ \hline
96 & 5 & 200 & 501 & 1224 &  $237.40 \pm 1.02$ \\
48 & 5 & 200 & 501 & 1224 &  $121.28 \pm 0.61$\\
32 & 3 & 200 & 501 & 1224 &  $81.35 \pm 0.34$\\
16 & 3 & 200 & 501 & 1224 &  $40.94 \pm 0.04$\\
\hline
\hline
\\[-0.75em]
\multicolumn{6}{c}{Matter} \\
$l_0$ & $n_{\text{sim}}$ & $\tau_\text{damp}$ & $\tau_\text{cg}$ & $\tau_{\text{end}}$ & $\xi_N(\tau = 0)$ \\ \hline
48 & 5 & 200 & 501 & 1224 & $120.51 \pm 0.11$ \\
\end{tabular}
\end{ruledtabular}

\caption{List of simulation parameters. Listed are: initial separation parameter $l_0$, number of simulations performed for each parameter set $n_\text{sim}$, time at end of damping $\tau_\text{damp}$, time at the end of core growth $\tauCG$, time at the end of simulation $\tau_\text{end}$. Additionally the mean separation of monopoles $\xi_N$ (see Eq.~(\ref{eq:xi_n})) immediately after initialisation is shown.}
\label{tab:param}
\end{table}

The covariant energy conservation was within 0.14\% for all runs, except for runs where $l_0 = 96$, which were within 0.6\%. The root mean square per-site Gauss law violation $\bar{G}/\bar{\rho}$ was less than $5 \times 10^{-4}$ for radiation era runs, and less than $10^{-3}$ for matter era runs, with $G$ and $\rho$ as given in Eqs.~(A12) and (A13) of Ref.~\cite{Hindmarsh:2016dha}.

\section{Measurements}
\label{sec:measure}

As we have seen in Sec.~\ref{sec:expectations}, defining a length scale and RMS velocity is of great importance to describe the large-scale motion of monopole networks. The combined analysis of these quantities is essential to correctly track the state of the system and consequently analyse the scaling behaviour. In this section we will show different ways to compute the network length scale and monopole velocities.

\subsection{Network length scale}

For monopole networks, a correct characterisation for the network length scale can be obtained using the comoving monopole separation:
\begin{align}
\xi = (\mathcal{V}/N)^{\frac13} \label{eq:xi_n},
\end{align}
where N is the total number of monopoles in the simulation volume $\mathcal{V}$. There are several ways to compute the number of monopoles. 

A direct way to compute the monopole number is by directly obtaining the topological charge in each lattice cell of the simulation box \cite{Lopez-Eiguren:2016jsy,Antunes:2002ss}, and counting the number of cells in which the topological charge is one or bigger. In this case we will have a monopole number, $N$, from directly counting them.  This length scale is denoted as $\xi_N$.

A way to reduce the computational cost of the length scale computation, taking into account that energy densities will be computed anyways, is the use of local field estimators.

The number of monopoles can be estimated through the formula (c.f. Refs.~\cite{Daverio:2015nva,Lopez-Eiguren:2017dmc})
\begin{equation}
   N_{E_V} = \frac{\tilde{E}_V}{a^s \tilde{M}_V} \equiv \frac{\sum_{\text{s}} d^3 x \; T_{00}(x,t) (V(\Phi)- V_0)}{a^s \sum_{\text{m}} d^3 x \; T_{00}(x,t) (V(\Phi) - V_0)}
   \label{eq:nev}
\end{equation}
where
\begin{equation}
   V_0 = \frac{1}{a^{2(1-s)}} \frac{3 m_0^4}{4\lambda_0}\label{eq:V0}
\end{equation}
is the minimum of the potential. The numerator is evaluated during our simulations, while the denominator comes from considering an isolated static monopole produced using gradient flow in a small box with twisted boundary conditions~\cite{Rajantie:2011nq}. The subscript `s' refers to an integration over the simulation box and `m' to a box in which the isolated monopole is present. Extrapolating our gradient flow computations to the infinite volume limit\footnote{The finite-volume effects go as $1/L$ for the mass, and are exponential for the potential-weighted `mass'.}, for our choice of mass and coupling parameters, we find the monopole mass $\mMon \approx 8.50$. Using Eq.~(\ref{eq:monopolemass}), this can be compared with the continuum numerical result. Interpolating the tabulated results in Ref.~\cite{Forgacs:2005vx} for our parameters ($m_\mathrm{H}/m_\mathrm{W} = g\sqrt{2\lambda} \approx 0.8944\ldots$), we obtain $\mMon \approx 8.58$, which is in good agreement with our lattice gradient flow computation.

Our lattice gradient flow calculation also yields the potential-weighted `mass' $\tilde{M}_V \approx 0.019$. Finally, noting that the monopole mass scales as $g^{-1}$ (see Equation~\ref{eq:monopolemass}), to account for expansion, we multiply the denominator by $a g/{g_0} = a^s$.

When $N_{E_V}$ is used to estimate the number of monopoles, we will denote the length scale as $\xi_{E_V}$.

The procedure performed in the integrals of Eq.~(\ref{eq:nev}) is known as weighting~\cite{Lopez-Eiguren:2016jsy}. The aim of weighting is to pinpoint regions of monopoles, and therefore, we should use a quantity that has a maximum at the monopoles and vanishes away from them. In the core of the monopoles the potential is out of its minimum and therefore, monopoles can be understood as concentrations of potential energy. This is the reason why we have chosen the potential as the weighting field. Moreover, potential weighting has been tested and verified as a good weighting quantity for monopoles in \cite{Lopez-Eiguren:2016jsy}.

The monopole density fraction defined in Eq.~(\ref{eq:densityfraction}) can be written using the network length scale $\xi$ defined in Eq.~(\ref{eq:xi_n}).  In radiation era it is
\begin{equation}
\Omega_\text{m}  = \frac{\rhom}{\rho_c} \propto \frac{a}{\xi^3}, \label{eq:measurement-densityfraction}
\end{equation}
and in matter era, where $a \propto \tau^2$, 
\begin{equation}
\Omega_\text{m}  = \frac{\rhom}{\rho_c} \propto \frac{1}{\xi^3} . \label{eq:measurement-densityfraction-matter}
\end{equation}

\subsection{Monopole velocity}

Similarly to the length scale case, monopole velocities can be computed from local field values. More precisely, velocity estimators can be derived from the Lorentz translation properties of the fields. In this subsection we will obtain those estimators from Lorentz boosted monopole fields, more detailed analysis can be found in \cite{seppa2023lattice}.

In order to compute velocity estimators from expressions of the Lorentz boost we will consider that all the energy in the field is in the form of monopoles. We will refer to local rest frame coordinates with $\mathbf{X}_\text{m}$ and generally use the subscript or superscript m to refer to the rest frame of the monopole. In order to simplify the procedure, the calculations of the velocity estimators will be performed in a Minkowski space-time. If we want to extrapolate the results to an expanding background it is enough to consider the space-time coordinates as comoving and the time coordinate as conformal. Physical quantities can be obtained multiplying by the scale factor $a$. The fields related to a moving monopole can be written in the rest frame as follows:
\begin{align}
\vb{E}(\vb{x},t) &= \gamma \dot{\vb{X}} \times \vb{B}_\text{m}(\vb{x}_\text{m}),\label{lorentz-E}\\
\vb{B}(\vb{x},t) &= \hat{\vb{v}}(\hat{\vb{v}} \cdot \vb{B}_\text{m}(\vb{x}_\text{m})) + \gamma\vb{B}_\text{m}^{\bot}(\vb{x}_\text{m}), \label{lorentz-B}\\
\pi^a(\vb{x},t) &= \gamma \dot{\vb{X}} \cdot \vb{D}\phi^a_\text{m}(\vb{x}_\text{m}),\label{lorentz-pi}\\
\vb{D}\phi^a(\vb{x},t) &= \hat{\vb{v}}(\hat{\vb{v}} \cdot \vb{D}\phi^a_\text{m}(\vb{x}_\text{m})) + \vb{D}^{\bot}\phi^a_\text{m}(\vb{x}),\label{lorentz-Dphi}
\end{align}
where $\dot{\vb{X}}$ is the velocity of the moving monopole,  $\gamma = 1/\sqrt{1-\dot{\vb{X}}^2}$ is the boost factor, and $\hat{\vb{v}}$ is the unit vector of the velocity. 

As we have seen in the previous subsection, in order to compute monopole related quantities it is useful to consider weighted variables. The weighted energy  and the Lagrangian of the system can be written as:
\begin{align}
E&=E_{\pi}+E_{D}+E_E+E_B+E_V, \label{eq:ebycomp}\\
L &=E_E-E_B+E_{\pi}-E_D-E_V,\label{eq:lbycomp}
\end{align} 
where each of the terms are defined as follows: 
\begin{align}
E_{\pi} &= \frac{1}{2} \int d^3x (\pi^a\pi^a )(V(\Phi)-V_0),\\
\label{eq:energies_pi}
E_D &= \frac{1}{2} \int d^3x ({\bf D} \phi^a \cdot {\bf D} \phi^a) (V(\Phi)-V_0),\\
E_E &= \int d^3x \tr( (\gamma \dot{\vb{X}} \times \vb{B}_\text{m}(\vb{x}))^2 )(V(\Phi)-V_0),\\
E_B &= \int d^3x \tr(\vb{B}^2)(V(\Phi)-V_0),\\
E_V &= \int d^3x V(\Phi)(V(\Phi)-V_0)).
\label{eq:energies_V}
\end{align}
where $V_0$ is defined in Eq.~(\ref{eq:V0}).

Making use of the redefinitions in Appendix~\ref{ap:lorentz} the energies can be rewritten as follows:
\begin{align}
E_{\pi} &= \frac13 \gamma v^2 E_D^\text{m},\\
E_D &= \frac{1}{3\gamma} (\gamma^2 + 2) E_D^\text{m},\\
E_E &= \frac{2}{3} \gamma v^2 E_B^\text{m},\\
E_B &= \frac{1}{3\gamma}(2\gamma^2 +1)E_B^\text{m}, \\ 
E_V &= \frac{1}{\gamma} E_V^\text{m}.
\end{align}

Analysing energy components, one can see that there are different ways to extract the monopole velocity $v$. First of all, it is clear that $E_E$ and $E_B$ are expressed in terms of two variables only, the monopole velocity v and $E_B^\text{m}$. Therefore, we can construct a velocity estimator from the ratio of those two energies:
\begin{equation}
      \label{eq:v_g}
\bar{v}^2_g=\frac{3 R_g}{2+R_g},
\end{equation}
where $R_g=E_E/E_B$ is the ratio of energies.

Similarly, $E_{\pi}$ and $E_D$ are functions of $E_D^\text{m}$ and $v$ only so denoting the velocity estimator arising from the scalar fields as $v_s$ we have,
\begin{equation}
   \label{eq:v_s}
\overline{v}_s^2 =  \frac{3R_s}{1+2R_s},
\end{equation}
where $R_s=E_{\pi}/E_D$.

Finally, we can rewrite the energy expression (\ref{eq:ebycomp}) and the Lagrangian expression (\ref{eq:lbycomp}) using energy components rewritten in rest frame variables:
\begin{align}
E &= \gamma(E_B^\text{m} + E_D^\text{m} + E_V^\text{m}) +\frac13\gamma v^2(E_B^\text{m} - E_D^\text{m}) \nn
&= \gamma \mMon. \label{eq:Erest}\\
L &= E_E - E_B + E_{\pi} - E_D - E_V \nn
&= -\frac{1}{\gamma} (E_B^\text{m} + E_D^\text{m} + E_V^\text{m}) \nn
&= -\frac{1}{\gamma} \mMon.  \label{eq:Lrest}
\end{align}

It is clear from Eqs.~(\ref{eq:Erest}-\ref{eq:Lrest})  that we can obtain another velocity estimator, which estimates the average velocity of monopoles using the Lagrangian density of the system,  hence we will denote it as $v_L$,
\begin{equation}
\overline{v}^2_{\mathcal{L}} = 1 + \frac{L}{E}.
   \label{eq:v_L}
\end{equation}

\section{Results \label{sec:res}}

\begin{figure} [t]
   \centering \includegraphics{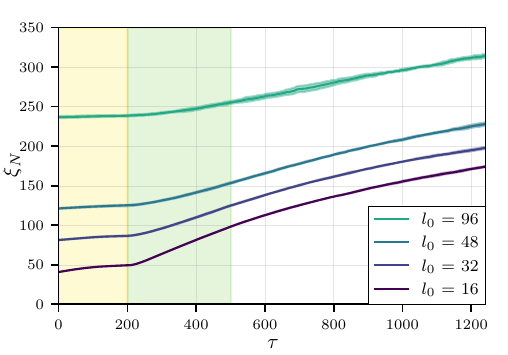}
   \caption{Mean monopole separation $\xi_N$ during the radiation era shown for different parameter choices $l_0$, plotted against conformal time $\tau$. Damping phase 0 - 200 $\tau$ shown in yellow, and growth phase 201-501 $\tau$ shown in light green. Errors obtained by averaging over the $n_\text{sim}$ simulations per parameter set.}
   \label{fig:xi_l0s}
\end{figure}

\begin{figure}  [t]
   \centering \includegraphics{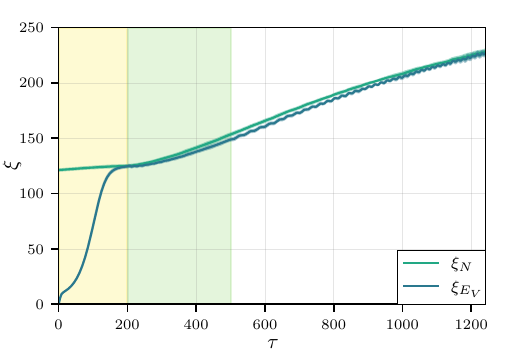}   
    \includegraphics{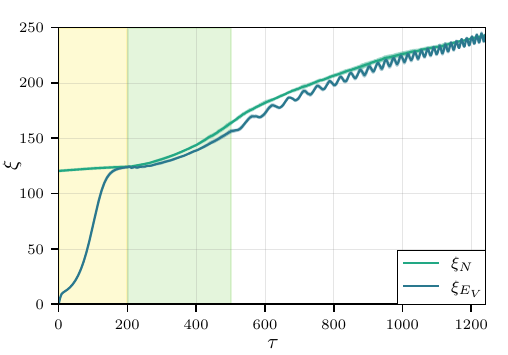}
   \caption{The two monopole separation estimators $\xi_N$ and $\xi_{E_V}$ plotted against conformal time $\tau$ for parameter choice $l_0 = 48$. Damping phase 0 - 200 $\tau$ shown in yellow, and growth phase 201-501 $\tau$ shown in light green. Top: radiation era. Bottom: matter era.}
   \label{fig:xi_l0_48}
\end{figure}

\subsection{Length Scales}

Measurements of the length scale in networks of topological defects are key to understanding whether the system exhibits scaling. Furthermore, the measurements can be used to determining whether the monopoles come to dominate the energy density.

The way in which we set the initial field configuration gives us good control over the initial monopole separation, with the mean monopole separation after initialisation being approximately $5/2~l_0$. In Fig.~(\ref{fig:xi_l0s}) we show the dependence of mean monopole separation $\xi_N$, defined in Eq.~(\ref{eq:xi_n}), on the parameter $l_0$ during the radiation era. During the main simulation phase, from $600~\tau$ onwards, all cases except for $l_0 = 96$ show similar growth with time, indicating scaling.

We have two monopole separation estimators at hand, $\xi_N$ and $\xi_{E_V}$, which we compare as a cross check. In Fig.~(\ref{fig:xi_l0_48}) these are shown for $l_0 = 48$ during radiation domination (top) and matter domination (bottom). We observe that the field energy based estimator $\xi_{E_V}$ has coherent oscillations which are clearly visible when entering the main simulation phase. We speculate that these oscillations result from the monopole core growth phase, and could potentially be lessened by slower core growth. Similar effect has been observed in string simulations, e.g. in Ref.~\cite{Hindmarsh:2018wkp}.
Overall, the two separation estimators agree well during both the growth phase and the main simulation phase. In the following, we will consider only $\xi_N$, which does not suffer from oscillations.

Rather than fitting $\xi_N$, we find it more instructive to consider the density fraction of monopoles, which we do in the next section.

\subsection{Density fraction}

\begin{figure} [t]
   \centering \includegraphics{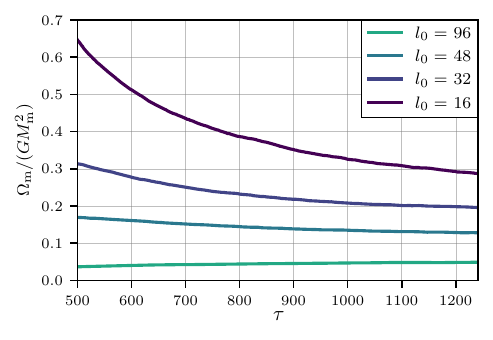}
   \includegraphics{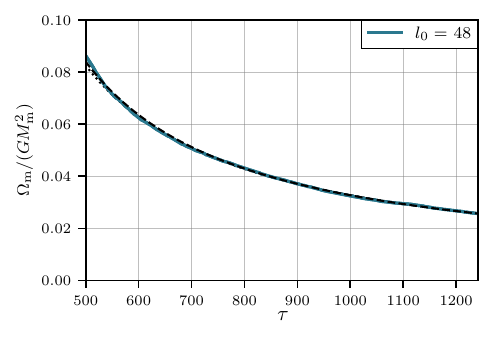}
   \caption{Top: The monopole density fraction during radiation era. Bottom: Matter era, with fit on Eq.~(\ref{eq:densfrac_mat_fit}).  Dashed lines are for fits from $\tau = 612$ onwards, dotted lines for fits from $\tau = 812$ onwards. For both, only the main simulation phase is shown.}
   \label{fig:dens_fraction}
\end{figure}

The density fraction can be estimated directly in terms of time series of $\xi_N$ obtained from simulations, using Eqs.~(\ref{eq:measurement-densityfraction}) and (\ref{eq:measurement-densityfraction-matter}). A plot of the density fraction measured this way is shown in Figure~(\ref{fig:dens_fraction}) for both matter and radiation epochs. 

In matter era we expect the asymptotic behaviour of the density fraction to follow Eq. (\ref{eq:densityfraction}). Due to this, we fit a function of form
\begin{align}
\dfrac{\Omega_\text{m}(\tau)}{GM_\text{m}^2} = \dfrac{8\pi}{3\nu^2} \frac{1}{M_\text{m}} \frac{\tau^2 a} { \xi^{3} }  = c_0 \dfrac{1}{\log^3{(\tau/c_1)}}  \label{eq:densfrac_mat_fit}
\end{align} 
to our $\xi_N$ data. To explore the scaling, we use fits on two different time ranges, $\tau \in [612,1223]$ and $\tau \in [812,1223]$. The longer range then goes from $1/8$ light-crossing time to half light-crossing time, with the damping phase excluded. 

The fitting parameters are shown in table (\ref{tab:fits_density}), and fit in Figure~(\ref{fig:dens_fraction}). Errors are estimated by jackknifing the 5 simulation realisations we have for matter era. In practice the fits are performed on a time series obtained by averaging over the simulation realisations with one removed. Fitting results and uncertainties are obtained by taking the average and standard error of the resulting fits.

\begin{table}[!h]
\begin{ruledtabular}
\begin{tabular}{  c  c  c  c  c  }
\multicolumn{5}{c}{Matter} \\
$\Delta \tau_{\textrm{fit}}$ & $l_0$ & $c_0$ & $c_1$ &  \\
\hline
612 - 1223 & 48 & 0.59   $\pm$    0.03  &  73.15   $\pm$      2.22 & \\
812 - 1223 & 48 & 0.56   $\pm$    0.05  &  76.47   $\pm$      5.38 & \\
\end{tabular}

\end{ruledtabular}

\caption{Numerical values of fit parameters from Eq.~(\ref{eq:densfrac_mat_fit}). Fits of density fraction on two ranges of conformal time. Values and errors are obtained by jackknifing as described in the text. } 
\label{tab:fits_density}
\end{table}

In radiation era, for the cases with fewer monopoles, ($l_0 = 48,~96$), there is clear indication of the density fraction tending towards a constant. The dynamic range of our simulation is not long enough to determine whether all cases reach an asymptotic density fraction $\Omega_\text{m}^{\infty}$. However, we observe that $l_0 = 96$ approaches a constant density fraction from below, while $l_0 = 48$ does so from above. Thus we postulate that the asymptotic density fraction is bracketed by the average density fraction at the end of simulation for $l_0 = 96$ and for $l_0 = 48$, which gives
\begin{equation}
0.049  \lesssim \Omega_\text{m}^{\infty} / (GM_\text{m}^2) \lesssim 0.128.
\label{e:RadEraDenFra}
\end{equation}

\subsection{Velocities}

\begin{figure}[h!]
   \centering \includegraphics{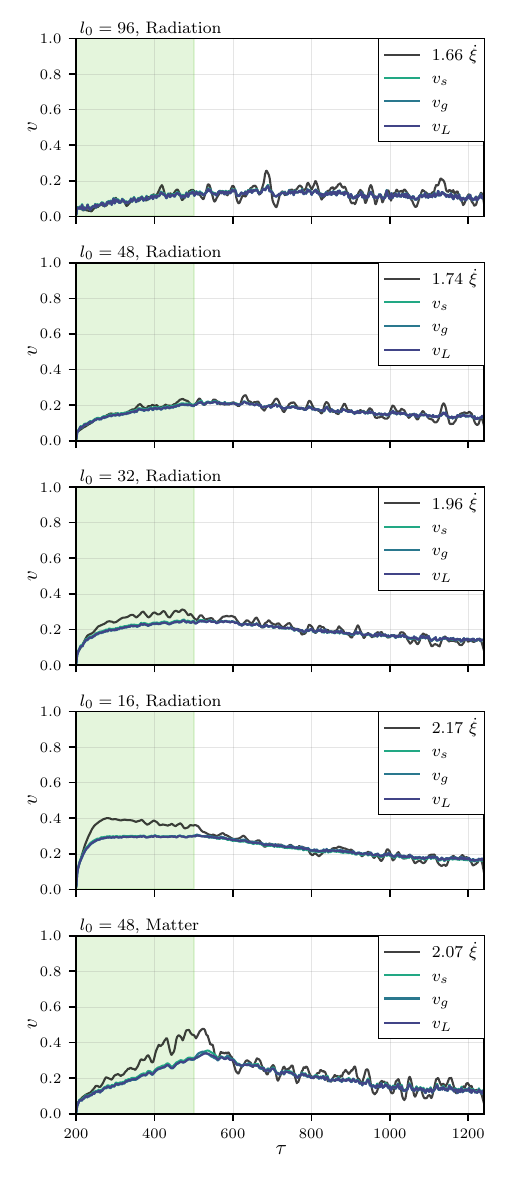}
   \caption{   
   Monopole velocity estimators and $d\xi_N/d\tau = \dot{\xi}$ compared during growth phase (201-501 $\tau$) and main simulation phase. Savitzky-Golay filter has been applied to $d\xi_N/d\tau$ to smooth the data. From top to bottom, $l_0 = 96, 48, 32, 16$ are shown for radiation era. Last (bottom) plot shows matter era.}
   \label{fig:velo_xi_comparison}
\end{figure}

To determine whether relation Eq.~(\ref{e:NewVOSxi}) holds, we plot the velocity estimators defined in Eqs. (\ref{eq:v_s}, \ref{eq:v_g} and \ref{eq:v_L}) in Fig.~(\ref{fig:velo_xi_comparison}), where we also show $d\xi_N/d\tau$. We obtain $d\xi_N/d\tau$ from $\xi_N$ via derivation. Savitzky-Golay filter with window length 21 and polynomial order 2 is applied to reduce the fluctuations arising from numerical derivation \cite{savitzky1964smoothing}. The different velocity estimators agree very closely both in growth and main simulation phase. 

For all our choices of initial separation, we see evidence of proportionality between $d\xi_N/d\tau$ and the monopole root mean square velocity. The proportionality constant was obtained as a mean of the ratio of $v_s$ and $d\xi_N/d\tau$ during main simulation phase. The proportionality constant gives $d$ of Eq. \ref{e:NewVOSxi}, with $d \approx 3/2$. The value of $d$, which parametrises the annihilation efficiency, is remarkably almost equal in matter and radiation epochs.

\subsection{Dynamical system for monopole evolution}
\label{s:DynSys}

We now go on to investigate the relationship between length scale and velocity, plotting the evolution of our system using the variables $(x,y)$ derived in Section~\ref{sec:modelling_xi}. We plot $(x,y)$ averaged over the $n_{\text{sim}}$ simulation realisations per parameter set.

\begin{figure}[tb]
   \centering \includegraphics[width=\linewidth]{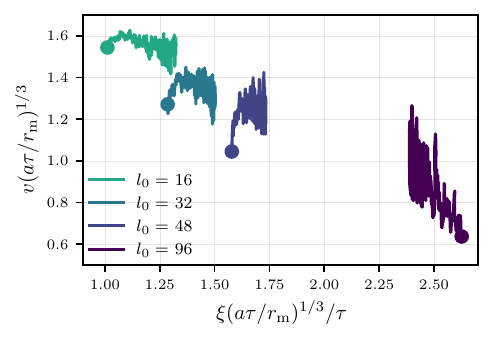} \\
   \includegraphics[width=\linewidth]{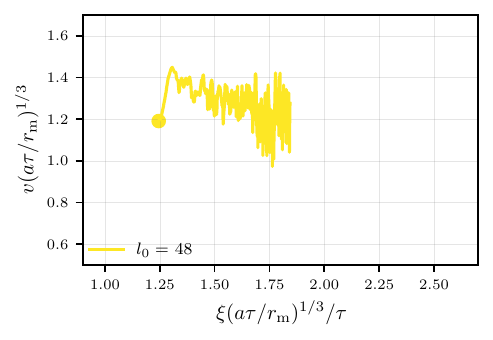}
   \caption{Plot of average $x, y$ for $\alpha=0$. Dot marks the start of main simulation phase ($\tau$ = 501). The $x,y$ are obtained by averaging over simulation realisations. Top: Radiation era. Bottom: Matter era.}
   \label{fig:phase_space}
\end{figure}

We start by presenting results for both our radiation and matter era simulations assuming $\alpha=0$, in Figure~\ref{fig:phase_space}. For the radiation era case, 
the evolution is consistent with a fixed point around $(x_*, y_*) \approxeq (2,1.2)$, with $l_0 = 96$ case approaching from right and the rest from left. 

For the matter era, our model predicts a fixed point at $x_* = \infty$ for any value of $y_*$. While it is hard to draw conclusions from one matter era simulation, there is some tentative indication that this might be the case, as the $x$ continually increases while $y$ remains approximately constant.

\begin{figure}[tb]
   \includegraphics[width=\linewidth]{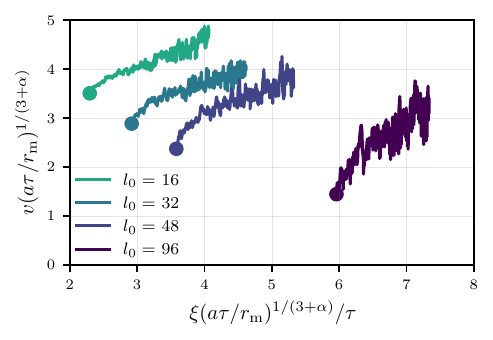}
   \caption{Plot of average $x, y$ for $\al = -1$ in Radiation era. Dot marks the start of main simulation phase ($\tau$ = 501).}  \label{fig:phase_space_alpha-minus1}
\end{figure}

\begin{figure}[tb]
   \includegraphics[width=\linewidth]{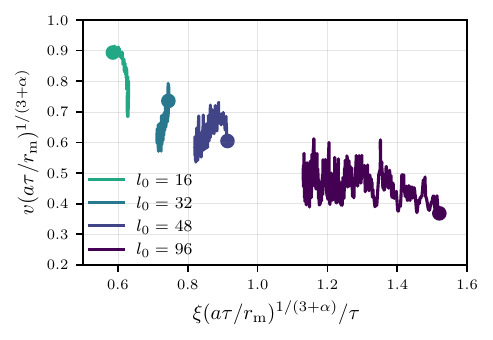}
   \caption{Plot of average $x, y$ for $\al = 3/2$ in Radiation era. Dot marks the start of main simulation phase ($\tau$ = 501).}  \label{fig:phase_space_alpha-32}
\end{figure}

In Fig.~\ref{fig:phase_space_alpha-minus1} we show phase space plots for $x_\al, y_\al$ with $\al = -1$, the minimum value proposed in Ref.~\cite{Martins:2008zz}.  We see that there is no evidence of a fixed point in the radiation era in these variables, as both continually increase for all initial conditions.

Finally, in Fig.~\ref{fig:phase_space_alpha-32}, we present a phase space plot for our radiation era results taking $\al = 3/2$, the maximum value proposed in Ref.~\cite{Sousa:2017wvx}. There is some evidence for a fixed point in this case, although not necessarily better than the $\al=0$ case presented above.

We thus conclude that there is some evidence for a fixed point $(x_*,y_*)$ for our radiation era simulations, with $\al = 0$ qualitatively more persuasive than $\al =3/2$. To differentiate between these models more conclusively, the dynamical range of the simulations would need to be extended.

\section{Discussion and conclusions \label{sec:dis}}

In this paper, we have carried out the first simulations of the dynamics of a gas of 't Hooft-Polyakov monopoles in the early universe, with the aim of testing velocity one-scale (VOS) models of their dynamics, in the simplified case when plasma interactions can be neglected.

Two such models have been proposed~\cite{Martins:2008zz,Sousa:2017wvx}, differing in their assumptions about how the Coulomb attraction operates in a monopole gas in an expanding universe. The modification to the Coulomb force term for mean comoving separation between monopoles $\xi$ and comoving horizon distance $\tau $ is multiplied by  $(\tau/\xi)^\alpha$ (Eq.~\ref{eq:ma_vos_vel}), with $\alpha = 0$ corresponding to no modification.  One model \cite{Martins:2008zz} argues that $\alpha = -1$, while the other \cite{Sousa:2017wvx} argues that $0 \le \alpha \le 3/2$. 

There is an explicit length scale in the VOS models,
proportional to the monopole radius $\rMon$. We show that the VOS model equations for the mean monopole separation and the RMS velocity can be recast in terms of an autonomous system in dimensionless variables ($x_\alpha$, $y_\alpha$), related to the originals by  scaling with $\alpha$-dependent powers of $\rMon$ and the cosmic time. The dynamical equations have a unique fixed point in the radiation era, and in the matter era a line of fixed points at zero monopole density, approached logarithmically in time.

Analysing our data with $\alpha = 0$ we find the predicted constant monopole density fraction in the radiation era, and in the matter era the density fraction decreases as $1/\ln^3(\mMon t)$, also as predicted.
Analysing our radiation era data with $\alpha = -1$~\cite{Martins:2008zz}, we find no evidence for a fixed point, which is inconsistent with the VOS model predictions at that parameter value.
Analysing our radiation era data with $\alpha = 3/2 $~\cite{Sousa:2017wvx}, we find marginal evidence for a fixed point: the time derivatives of the mean separation variable $x_\alpha$ are all negative at the end of the simulation, but the magnitude $|\tau \dot x_\al | $ is very small, and it could be argued that there is a fixed point at $(x_{\al *}, v_{\al*}) \simeq (0.6, 0.5)$.

We find the form for the screening effect proposed in Refs.~\cite{Martins:2008zz,Sousa:2017wvx} difficult to understand. It proposes that the screening effect operates by changing the power law of the Coulomb force, while screening in more familiar thermal plasmas makes the force decay exponentially at large distances beyond the Debye length $\la_D$, which depends on temperature and charge carrier density. Furthermore, the length scale proposed in Refs.~\cite{Martins:2008zz,Sousa:2017wvx} is the horizon length, for all monopole separations, which should not be a relevant length scale for monopole separations much less than the horizon. We have seen that at a fixed point the monopole separation increases less quickly than the horizon distance, and so the mean monopole separation is always much less than the horizon distance.  While a screening effect can be expected, the only relevant length scale in the problem is the mean monopole separation, which is
implicit in the model assumption that the RMS acceleration is proportional to the inverse square of the separation.

In summary, our results are consistent with the VOS model with the theoretically well-motivated value $\al=0$. This predicts that the radiation era monopole density fraction (in the absence of thermal effects) is proportional to $G\mMon^2$, while the matter era monopole density decreases as $G\mMon^2/\ln^3(\mMon t)$.  Our radiation era measurements are consistent with a constant in a range given in Eq.~\eqref{e:RadEraDenFra}, and our matter era measurements give a very good fit to the predicted functional form (see Fig.~\ref{fig:dens_fraction} and Table~\ref{tab:fits_density}).

Our simulations were principally designed to test the VOS models, but as we have pointed out, 
in the early universe the monopoles interact with a plasma, which includes the quanta of its gauge field, principally the massless bosons of the unbroken gauge symmetry.  
The interaction would give them thermal kinetic energy, which drops less quickly than the average binding energy, and the annihilation of monopoles is predicted to eventually halt~\cite{Zeldovich:1978wj, Preskill:1979zi}.

We leave the examination of the dynamics of a thermal monopole gas to a future work.
By including thermal fluctuations, one could also check whether the gauge monopole formation mechanism proposed in Ref.~\cite{Rajantie:2002dw} affects the late-time dynamics. 
Even without thermal fluctuations, our simulation framework could be used to study the sweeping up of monopoles by domain walls~\cite{Dvali:1997sa,Bachmaier:2023zmq} in an expanding universe, with a suitable extension of the field content and symmetries.

\begin{acknowledgments}
The simulations for this paper were carried out at the
Finnish Centre for Scientific Computing CSC.  We wish to thank Lara Sousa and Ana Achúcarro for useful discussions.~MH (ORCID 0000-0002-9307-437X) was supported by Academy of Finland grant 333609, Research Council of Finland grant 363676 and STFC grant ST/X000796/1.
ALE (ORCID ID0000-0002-1696-3579)  was supported by Eusko Jaurlaritza IT1628-22 and by the PID2024-156016NB-I00 grant funded by MCIN/AEI/10.13039/501100011033/ and by ERDF; “A way of making Europe”.
RS (ORCID 0000-0002-1461-2644) was supported by a working grant from the Magnus Ehrnrooth foundation and Research Council of Finland grant 349865.
DJW (ORCID 0000-0001-6986-0517) was supported by Research Council of Finland grant nos. 324882, 328958, 349865 and 353131.
\end{acknowledgments}

\section*{Data Access Statement}

The simulation time series data is available at \url{https://doi.org/10.5281/zenodo.17632300}.

\appendix

\section{Monopole dynamics in FLRW \label{app:mono_dynamics}}

We start with the action of a particle with mass $\mMon$ and charge $\gm$ moving on a path $X^\mu(\lambda)$ in a spacetime with metric $g_{\mu\nu}$, 
\begin{eqnarray}
S &=& \int d^4x \sqrt{-g} \left( - \frac{1}{4} F_{\mu\nu} F^{\mu\nu} + A_\mu J^\mu\right)  \nonumber\\
&& - \mMon \int d\lambda \sqrt{\left(\frac{dX}{d\lambda}\right)^2}
\end{eqnarray}
where
\begin{equation}
J^\mu = \gm \int d\lambda \frac{d X^\mu}{d\lambda} (-g)^{-1/2} \delta^{(4)}(x - X(\la))
\end{equation}
is the current. If the particle is a magnetically charged, we interpret $A_\mu$ as the dual gauge field, and $F_{\mu\nu}$ as the dual field strength tensor, with $F_{0i} = B_i$.
This action gives as equations of motion 
\begin{equation}
\partial_\nu \left( \sqrt{-g} F^{\mu\nu} \right) = \gm \int d\lambda \frac{d X^\mu}{d\lambda} \delta^{(4)}(x - X(\la)).
\end{equation}
and
\begin{equation}
 \frac{d^2 X^\mu}{d\lambda^2} + {\Ga^\mu}_{\nu\rh}  \frac{d X^\nu}{d\lambda}  \frac{d X^\rh}{d\lambda} 
 =
 \gm {F^\mu}_\nu  \frac{d X^\nu}{d\lambda}.
\end{equation}
In a FLRW background in conformal coordinates, the metric is 
\begin{equation}
g_{\mu\nu} = a^2(\tau) \eta_{\mu\nu}. 
\end{equation} 
the equation of motion for a monopole is 
\begin{equation}
\frac{d^2 X^i}{d\lambda^2} + 2 \frac{\dot a}{a} \frac{d X^i}{d\lambda}  = \gm {F^i}_0 \frac{d X^0}{d\lambda} .
\end{equation}
Using Gauss's law  for the magnetic field, 
\begin{equation}
\pa_i F_{0i} =  \gm  \delta^{(3)}(\bx - \bX(\ta))
\end{equation}
where $\ta \equiv X^0$, the magnetic field is 
\begin{equation}
F_{0i}(\bx) = \frac{\gm}{4\pi} \frac{\hat{X}_i}{|\bx - \bX(\ta)|^2}. 
\end{equation}
In the non-relativistic approximation,
\begin{equation}
\frac{d X^0}{d\lambda}  = \frac{1}{a},
\end{equation}
and hence the equation of motion for a monopole moving in the field of an antimonopole fixed at the origin is 
\begin{equation}
\frac{d^2 \bX}{d\tau^2} + \frac{\dot a}{a} \frac{d \bX}{d\tau}  = \frac{\gm^2}{4\pi} \frac{\hat\bX}{a |\bX|^2} .
\end{equation}

\section{Lorentz Boost in Energy Components \label{ap:lorentz}}

Energies in (\ref{eq:energies_pi}-\ref{eq:energies_V}) can be rewritten using equations (\ref{lorentz-E}-\ref{lorentz-Dphi}) and transverse fields, which are defined through projection operators,
\begin{align}
D_i^{\bot} \phi^a(\vb{x},t) &= (\delta_{ij} - \hat{v}_i\hat{v}_j)D_j\phi^a(\vb{x},t), \\
B_i^{\bot}(\vb{x},t) &= (\delta_{ij} - \hat{v}_i\hat{v}_j)B_j(\vb{x},t).
\end{align}
With this redefinition, the energy components are described fully by local frame energies and velocity $v$:
\begin{align}
E_{\pi} &= \frac{1}{2} \int d^3x_\text{m} \gamma v^2 \hat{v}^i \hat{v}^j (D^i\phi^a_\text{m} D^j\phi^a_\text{m} ) = \frac13 \gamma v^2 E_D^\text{m}.\\
E_D &= \int d^3x_\text{m} \frac{1}{\gamma}(\gamma^2 \hat{v}^i\hat{v}^j + \delta^{ij} - \hat{v}^i\hat{v}^j )(D^i\phi^a_\text{m}D^j\phi^a_\text{m}) \nn
&= \frac{1}{3\gamma} (\gamma^2 + 2) E_D^\text{m}.\\
E_E &= \int d^3x_\text{m}  \frac{1}{\gamma} \tr( \gamma^2 \varepsilon^{ijk} v^j B_\text{m}^k \varepsilon^{iln} v^l B_\text{m}^n ). \label{eq:restEE}\\
E_B &= \int d^3x_\text{m} \frac{1}{\gamma} (\hat{v}^i\hat{v}^j + \gamma^2 (\delta^{ij} - \hat{v}^i\hat{v}^j) ) \tr(B_\text{m}^iB_\text{m}^j).\label{eq:restEB}\\
E_V &= \int d^3x V(\Phi) = \int d^3x_\text{m} \frac{1}{\gamma} V(\Phi_\text{m}) = \frac{1}{\gamma} E_V^\text{m}.
\end{align}
Working with the Levi-Civita identity and making use of the spherical symmetry Eqs.~(\ref{eq:restEE}) and (\ref{eq:restEB})  can be written as 
\begin{align}
E_E &= \int d^3x_\text{m} \gamma v^2 \tr(B_\text{m}^i B_\text{m}^i) - \int d^3x_\text{m} \frac13 \gamma v^2 \tr(B_\text{m}^i B_\text{m}^i) \nn
&= \frac{2}{3} \gamma v^2 E_B^\text{m}.\\
E_B &= \int d^3x_\text{m} \frac{1}{\gamma} (\hat{v}^i\hat{v}^j + \gamma^2 (\delta^{ij} - \hat{v}^i\hat{v}^j) ) \tr(B_\text{m}^iB_\text{m}^j)\nn
&= \frac{1}{3\gamma}(2\gamma^2 +1)E_B^\text{m}.
\end{align}

\bibliography{scone}

\end{document}